\documentclass[aps,prl,twocolumn,showpacs]{revtex4}

\usepackage{graphicx}
\usepackage{graphics}
\usepackage{amsfonts}
\usepackage{amsmath}
\usepackage{epsfig}

\begin{document}

\newcommand{\li}{$^6$Li}
\newcommand{\na}{$^{23}$Na}
\newcommand{\cs}{$^{133}$Cs}
\newcommand{\kk}{$^{40}$K}
\newcommand{\rb}{$^{87}$Rb}
\newcommand{\vect}[1]{\mathbf #1}
\newcommand{\g}{g^{(2)}}
\newcommand{\one}{$\left|1\right>$}
\newcommand{\two}{$\left|2\right>$}
\newcommand{\V}{V_{12}}
\newcommand{\kfa}{\frac{1}{k_F a}}

\title{Observation of Fermi Polarons in a Tunable Fermi Liquid of Ultracold Atoms}

\author{Andr\'e Schirotzek, Cheng-Hsun Wu, Ariel Sommer, and Martin W. Zwierlein}
\affiliation{Department of Physics, MIT-Harvard Center for Ultracold Atoms, and Research Laboratory of Electronics,
Massachusetts Institute of Technology, Cambridge, Massachusetts 02139, USA}

\begin{abstract}
We have observed Fermi polarons, dressed spin down impurities in a spin up Fermi sea of ultracold atoms.
The polaron manifests itself as a narrow peak in the impurities' rf spectrum that emerges from a broad incoherent background. We determine the polaron energy and the quasiparticle residue for various interaction strengths around a Feshbach resonance.
At a critical interaction, we observe the transition from polaronic to molecular binding.
Here, the imbalanced Fermi liquid undergoes a phase transition into a Bose liquid coexisting with a Fermi sea.
\end{abstract}

\maketitle

The fate of a single impurity interacting with its environment determines the low-temperature behavior of many condensed matter systems. A well-known example is given by an electron moving in a crystal lattice, displacing nearby ions and thus creating a localized polarization. The electron, together with its surrounding cloud of lattice distortions, phonons, forms the lattice polaron~\cite{land33polaron}. It is a quasiparticle with an energy and mass that differ from that of the bare electron. Polarons are central to the understanding of colossal magnetoresistance materials~\cite{mann05quasi}, and they affect the spectral function of cuprates, the parent material of High-$T_C$ superconductors \cite{lee06hightc}. Another famous impurity problem is the Kondo effect, where immobile spin impurities give rise to an enhanced resistance in metals below the Kondo temperature \cite{kond64}. In contrast to the electron moving in a phonon bath, a bosonic environment, in the latter case the impurity interacts with a fermionic environment, the Fermi sea of electrons.

\begin{figure}
\includegraphics[width=3in]{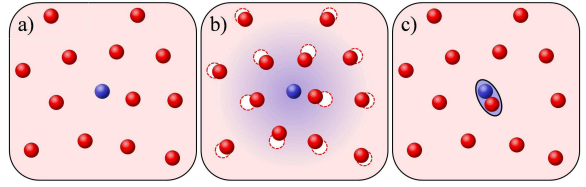}
\caption{From polarons to molecules. a) For weak attraction, an impurity (blue) experiences the mean field of the medium (red). b) For stronger attraction, the impurity surrounds itself with a localized cloud of environment atoms, forming a polaron. c) For strong attraction, molecules of size $a$ form despite Pauli blocking of momenta $\hbar k < \hbar k_F \ll
\hbar/a$ by the environment.}
\label{f:cartoon}
\end{figure}

Here we study a small concentration of spin down impurities immersed in a spin up Fermi sea of ultracold atoms. This system represents the limiting case of spin-imbalanced Fermi gases and has been recognized to hold the key to the quantitative understanding of the phase diagram of imbalanced Fermi mixtures~\cite{chev06polaron,lobo06polaron,schu07pair,bulg07asym,comb07polaron,
punk07rf,veil08RFimbal,pila08phase,prok08polaron,prok08boldpolaron,mass08polaron,comb08fullpolaron}.
Unlike in liquid $^3$He, the s-wave interaction potential between the impurities and the spin up atoms in this novel spin-imbalanced Fermi liquid is attractive. The vicinity of a Feshbach resonance allows to tune the interaction strength at will, characterized by the ratio of the interparticle distance $\sim 1/k_F$ to the scattering length $a$, where $k_F$ is the spin up Fermi wavevector~\cite{kett08varenna}. Fig.~\ref{f:cartoon} depicts the scenario for a single impurity:
For weak attraction ($1/k_F a \ll -1$) the impurity propagates freely in the spin up medium of density $n_\uparrow= k_F^3/6\pi^2$ (Fig.~\ref{f:cartoon}a). It merely experiences the familiar attractive mean field energy shift $E_\downarrow = 4\pi\hbar^2 a n_\uparrow / m < 0$.
However, as the attractive interaction grows, the impurity can undergo momentum changing collisions with environment atoms, and thus starts to attract its surroundings. The impurity ``dressed" with the localized cloud of scattered fermions constitutes the Fermi polaron (Fig.~\ref{f:cartoon}b). Dressing becomes important once the mean free path $\sim 1/n_\uparrow a^2$ of the bare impurity in the medium becomes comparable to the distance $\sim 1/k_F$ between environment particles or when $(k_F a)^2 \sim 1$. Collisions then reduce the bare impurity's probability of free propagation, the quasiparticle residue $Z$, from unity. The dressed impurity can instead move freely through the environment, with an energy $E_\downarrow$ shifted away from the simple mean field result.
This polaronic state is stable until, for strong attraction ($1/k_F a\sim 1$), equivalent to a deep effective potential well, the spin down impurity will bind exactly one spin up atom, thus forming a tightly bound molecule (Fig.~\ref{f:cartoon}c). This molecule is itself a dressed impurity, albeit a bosonic one~\cite{prok08polaron}.

\begin{figure*}
\includegraphics[width=5.5in]{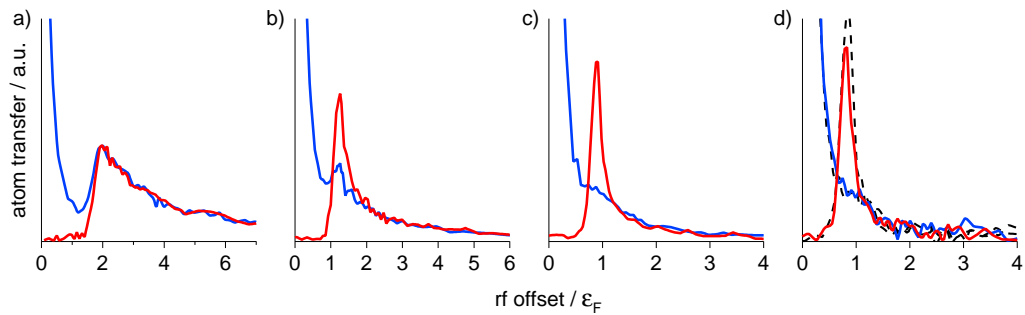}
\caption{Rf spectroscopy on polarons. Shown are spatially resolved, 3D reconstructed rf spectra of the environment (blue, state $\left|1\right>$) and impurity (red, state $\left|3\right>$) component in a highly imbalanced spin-mixture. a) Molecular limit, b), c) Emergence of the polaron, a distinct peak exclusively in the minority component. d) At unitarity, the polaron peak is the dominant feature in the impurity spectrum, which becomes even more pronounced for $1/k_F a < 0$ (not shown). For the spectra shown as dashed lines in d) the roles of states $\left|1\right>$ and $\left|3\right>$ are exchanged. The local impurity concentration was $x=5(2)\%$ for all spectra, the interaction strengths $1/k_F a$ were a) $0.76(2)$ b) $0.43(1)$ c) $0.20(1)$, d) $0$ (Unitarity).}
\label{f:spectra}
\end{figure*}

To prepare and observe Fermi polarons, we start with a spin-polarized cloud of $^6$Li atoms in the lowest hyperfine state $\left|1\right>$ (spin up), confined in a cylindrically symmetric optical trap ($125\, \rm \mu m$ waist, $145\,\rm Hz / 22.3\,\rm Hz$ radial/axial trapping frequency) at a magnetic field of 690 G~\cite{kett08varenna}. A two-photon Landau-Zener sweep transfers a small fraction into state $\left|3\right>$ (spin down), and further cooling results in a cloud containing 2\% $\left|3\right>$ impurities immersed in a degenerate Fermi gas of 5 million $\left|1\right>$ atoms at a temperature $T = 0.14(3) T_F$, where $T_F$ is the Fermi temperature. A 100 G wide Feshbach resonance for scattering between these states is centered at 690 G. For various fields around the resonance, we perform rf spectroscopy on the impurity species $\left|3\right>$ and on the environment particles in $\left|1\right>$ by transferring atoms into the empty state $\left|2\right>$, accessible to either hyperfine state. This state is sufficiently weakly interacting with the initial states to allow a direct interpretation of the resulting spectra~\cite{schu08pairsize}. As in previous work, spectra are spatially resolved and tomographically 3D reconstructed~\cite{shin07rf} via an inverse Abel transform, and are thus local and free from broadening due to density inhomogeneities. In addition, phase contrast images yield the in-situ density distribution $n_\uparrow$, $n_\downarrow$ and thus the local Fermi energy $\epsilon_F$ of the environment atoms and the local impurity concentration $x=\frac{n_{\downarrow}}{n_{\uparrow}}$. The Rabi frequencies $\Omega_R$ for the impurity and environment rf transitions are measured (on fully polarized samples) to be identical to within 5\%.

Fig.~\ref{f:spectra} shows the observed spectra of the spin down impurities and that of the spin up environment at low local impurity concentration. The bulk of the environment spectrum is found at zero offset, corresponding to the free (Zeeman plus hyperfine) energy splitting between states $\left|1\right>$ and $\left|2\right>$. However, interactions between impurity and spin up particles lead to a spectral contribution that is shifted: The rf photon must supply additional energy to transfer a particle out of its attractive environment into the final, non-interacting state~\cite{kett08varenna}. In Fig.~\ref{f:spectra}a), impurity and environment spectra above zero offset exactly overlap, signalling two-body molecular pairing. The steep threshold gives the binding energy, the high-frequency wings arise from molecule dissociation into remnants with non-zero momentum~\cite{kett08varenna,rega03mol,footpauli}.
As the attractive interaction is reduced, however, a narrow peak appears in the impurity spectrum that is not matched by the response of the environment (Fig.~\ref{f:spectra}b,c,d).
This narrow peak, emerging from a broad incoherent background, signals the formation of the Fermi polaron, a long-lived quasiparticle. The narrow width and long lifetime are expected: At zero temperature the zero momentum polaron has no phase space for decay and is stable. At finite kinetic energy or finite temperature $T$ it may decay into particle-hole excitations~\cite{prok08polaron}, but phase space restrictions due to the spin up Fermi sea and conservation laws imply a decay rate $\propto (T/T_F)^2\sim 1\%$ in units of the Fermi energy. Indeed, the width of the polaron peak is consistent with a delta function within the experimental resolution, as calibrated by the spectra of fully polarized clouds. The background is perfectly matched by the rf spectrum of the environment. This is expected at high rf energies $\hbar \omega \gg \epsilon_F$ that are probing high momenta $k \gg k_F$ and thus distances short compared to the interparticle spacing. Here, an impurity particle will interact with only one environment particle, leading to overlapping spectra.

F. Chevy has provided an instructive variational wavefunction~\cite{chev06polaron,comb07polaron} that captures the essential properties of the polaron, even on a quantitative level~\cite{comb08fullpolaron} when compared with Monte-Carlo (MC) calculations~\cite{lobo06polaron,prok08polaron,pila08phase}:

\begin{equation}
\left|\Psi\right> = \varphi_0 \left|\mathbf{0}\right>_\downarrow \left|FS\right>_\uparrow + \mathop{\sum_{\left|\mathbf{q}\right|<k_F<\left|\mathbf{k}\right|}} \varphi_{\mathbf{k}\mathbf{q}} c^\dagger_{\mathbf{k}\uparrow} c_{\mathbf{q}\uparrow}  \left|\mathbf{q}-\mathbf{k}\right>_\downarrow \left|FS\right>_\uparrow
\label{e:wavefunction}
\end{equation}

The first part describes a single impurity with a well-defined wavevector ($\vect{k}_\downarrow = \mathbf{0}$) that is not localized and free to propagate in the Fermi sea of up spins $\left|FS\right>_\uparrow$. In the second part the impurity particle recoils off environment particles that are scattered out of the Fermi sea and leave holes behind. This describes the dressing of the impurity with particle-hole excitations. The probability of free propagation is given by the first, unperturbed part, $Z = |\varphi_0|^2$.
According to Fermi's Golden Rule~\cite{kett08varenna,mass08polaron,veil08RFimbal,footepaps}, the two portions of $\left|\Psi\right>$ give rise to two distinct features of the impurity rf spectrum $\Gamma(\omega)$ ($\omega$ is the rf offset from the bare atomic transition):
\begin{eqnarray}
    \Gamma(\omega) = 2\pi\hbar\Omega_R^2\, Z \delta(\hbar\omega+E_\downarrow) + \Gamma^{\rm inc}(\omega)
    \label{e:spectrum}
\end{eqnarray}
The first part in $\left|\Psi\right>$ contributes a coherent narrow quasiparticle peak to the minority spectrum. Its position is a direct measure of the polaron energy $E_\downarrow$, its integral gives the quasiparticle residue $Z$.
The particle-hole excitations in the second part give rise to a broad, incoherent background $\Gamma^{\rm inc}(\omega)\propto \sum_{\mathbf{q},\mathbf{k}} \left|\varphi_{\mathbf{q}\mathbf{k}}\right|^2 \delta(\hbar\omega - \epsilon_{\mathbf{q}-\mathbf{k}} - \epsilon_\mathbf{k} + \epsilon_\mathbf{q} + E_\downarrow)$: The polaron energy $E_\downarrow$ is released as the impurity at momentum $\vect{q}-\vect{k}$ is transferred into the final state, leaving behind an environment particle in $\vect{k}$ above and a hole within the Fermi sea at $\vect{q}$~\cite{footepaps}.
These two spectral features are recovered in theoretical rf spectra for a finite number of impurities, i.e. a Fermi liquid~\cite{mass08polaron, schn09RF}. For our analysis we do not rely on a theoretical fit to the spectra.

\begin{figure}
\includegraphics[width=2.5in]{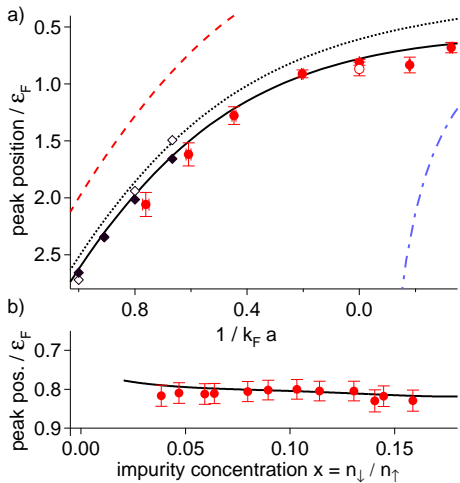}
\caption{Peak position of the impurity spectrum as a measure of the polaron energy $E_{\downarrow}$. a) peak position for various interaction strengths in the limit of low concentration $x=5(2)\%$ (solid circles). Open circle: Reversed roles of impurity and environment. Dotted line: polaron energy from variational Ansatz Eq.(\ref{e:wavefunction})~\cite{chev06polaron}, the solid line including weak final state interactions. Dashed line: Energy of a bare, isolated molecule in vacuum. Blue dash-dotted line: Mean field limit for the energy of an impurity atom. Solid (open) diamonds: Diagrammatic MC energy of the polaron (molecule)~\cite{prok08polaron}.
b) Peak position at unitarity ($1/k_F a=0$) as a function of impurity concentration (solid circles). The line shows the expected peak position, $\hbar\omega_p/\epsilon_F = A+(1-\frac{m}{m^*})x^{2/3}-\frac{6}{5}Fx+\frac{4}{3\pi}k_F a_{\rm fe}$, using the MC value $A = 0.615$~\cite{prok08polaron}, the analytic result $m^* = 1.2$~\cite{comb08fullpolaron}, the weak repulsion between polarons with $F=0.14$~\cite{pila08phase} and weak final state interactions with scattering length $a_{\rm fe}$.}
\label{f:maximaF}
\end{figure}

To measure the polaron energy $E_{\downarrow}$, we determine the peak position of the impurity spectrum as a function of the local interaction parameter $1/k_F a$.
The data for 5\% impurity concentration are shown in Fig.~\ref{f:maximaF}a), along with the variational upper bound given by the wavefunction Eq.(\ref{e:wavefunction})~\cite{footepaps} and the diagrammatic MC calculation of~\cite{prok08polaron}. As final state interactions are weak, they can be included as a simple repulsive mean field shift $4\pi\hbar^2 a_{\rm fe} n_\uparrow/m$, with $a_{\rm fe}$ the scattering length between the final state and the environment atoms~\cite{footfinalstate}.
Polaron energies have been predicted via the variational Ansatz~\cite{chev06polaron}, the T-matrix approach~\cite{comb07polaron,punk07rf,comb08fullpolaron,mass08polaron}, the $1/N$ expansion~\cite{veil08RFimbal}, fixed node MC~\cite{lobo06polaron,pila08phase} and diagrammatic MC~\cite{prok08polaron}. With the exception of the $1/N$ expansion, these all agree with each other and with the present experiment to within a few percent. In particular, in the unitary limit where $1/k_F a=0$ we find a polaron energy of $E_{\downarrow}=-0.64(7)\epsilon_F$ ($-0.72(9)\epsilon_F$) when state $\left|3\right>$ ($\left|1\right>$) serves as the impurity~\cite{footswitch}. This agrees well with the diagrammatic MC calculation, $-0.615\epsilon_F$~\cite{prok08boldpolaron}, and the analytical result $-0.6156(2)\epsilon_F$~\cite{comb08fullpolaron}. Analysis of experimental density profiles yields a value of $-0.58(5)\epsilon_F$~\cite{shin08EOS}.

The relatively large value for $E_\downarrow$ directly implies that the normal state, modeled as a Fermi sea of weakly interacting polarons, is favored over the superfluid state up to a critical concentration (44\%), much higher than that predicted by mean field theories (4\%)~\cite{shee06phase}. These neglect interactions in the normal state and therefore imply a polaron binding energy of zero.

We have so far considered the limit of few impurities. By increasing their density, we can study the effect of interactions between polarons. In Fig.~\ref{f:maximaF}b) we show that the quasiparticle peak position depends only weakly on the impurity concentration in the unitarity limit. Polarons are thus weakly interacting quasiparticles, despite the strong interactions between the bare impurity and its environment.

The peak position could be modified due to the effective mass $m^*$ of polarons, larger than the mass of the bare impurity. Transfer of a moving polaron into the free final state then requires additional kinetic energy. This leads to an upshift and a broadening on the order of the Fermi energy difference between initial and final state, $x^{2/3}\epsilon_F\left(1-\frac{m}{m^*}\right)$. On resonance, this is $0.04 \epsilon_F$ for $x = 0.1$. The effect could be partially masked by the predicted weak repulsion between polarons~\cite{pila08phase} that would downshift the resonance frequency by $-0.02 \epsilon_F$ for $x=0.1$.

\begin{figure}
\includegraphics[width=2.75in]{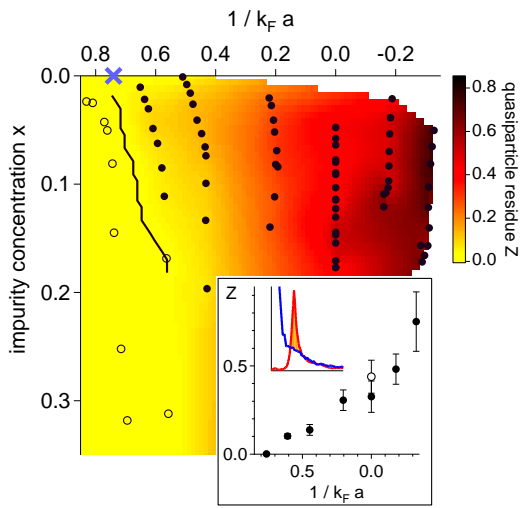}
\caption{Quasiparticle residue $Z$ as a function of interaction strength and impurity concentration. The color coding indicates the magnitude of $Z$ and is an interpolation of the data points shown in the graph. Open circles: Data points consistent with zero ($Z<0.03$), solid circles: $Z>0.03$, the solid line marking the onset of $Z$. Blue cross: Critical interaction strength for the Fermi liquid - molecular BEC transition for $x\rightarrow 0$~\cite{pila08phase}. Inset: $Z$ as a function of interaction strength in the limit of low impurity concentration $x=5(2)\%$. Open circle: Reversed roles, $\left|1\right>$ impurity in $\left|3\right>$ environment. The spectrum in the inset illustrates the determination of Z\cite{footepaps}.}
\label{f:residue}
\end{figure}

The spectral weight of the polaron peak directly gives the quasiparticle residue $Z$, a defining parameter of a Fermi liquid. Experimentally, we determine the area under the impurity peak that is not matched by the environment's response and divide by the total area under the impurity spectrum (see spectrum in Fig.~\ref{f:residue} and~\cite{footepaps}).
Fig.~\ref{f:residue} presents $Z$ as a function of interaction strength and impurity concentration $x$, the inset shows $Z$ for $x=5\%$. As expected, $Z$ approaches $100\%$ for weak attractive interaction $k_F a \rightarrow 0^-$, where the bare impurity only rarely recoils off environment atoms. As the mean free path shortens and the bare impurity starts to surround itself with environment atoms $Z$ decreases. On resonance, we find $Z = 0.39(9)$ for $x=5\%$, with only a weak dependence on $x$ (Fig.~\ref{f:residue}). Theoretical values for $Z$ vary: Ansatz Eq.(\ref{e:wavefunction}) predicts $Z = 0.78$ for a single impurity, while Ref.~\cite{veil08RFimbal} predicts $Z=0.47$ ($0.30$) for vanishing ($5\%$) impurity concentration.
Our procedure might yield a lower bound on the actual value of $Z$, as the incoherent part of the impurity spectrum might be depleted around threshold.
Eventually, for strong attraction between the impurity and particles of the medium, $Z$ vanishes and we observe complete overlap of the impurity and environment spectra. This signals the formation of a two-body bound state between the impurity and exactly one environment atom. % It is expected to occur once the size of free-space molecules $~a$ becomes smaller than the interparticle spacing $~1/k_F$, so that Pauli blocking at momenta $<\hbar k_F$ does no longer prevent formation of a bound state with momentum spread $~\hbar/a$.
For a spin down concentration of $5\%$ we determine the critical interaction strength where the polaron peak vanishes to be $1/(k_F a)_c = 0.76(2)$. This is in good agreement with the independently determined critical interaction $1/k_F a = 0.74(4)$ beyond which one finds a superfluid even for the smallest impurity concentration~\cite{shin08bosefermi}. This is a multicritical point~\cite{sach06lutt,pila08phase,prok08polaron} where a Fermi liquid of weakly interacting polarons undergoes a phase transition into a Bose liquid of molecular impurities. Fixed-node MC calculations place this transition at a value of $1/k_F a = 0.73$ for $x\rightarrow 0$~\cite{pila08phase}.
Our $1/(k_F a)_c$ is lower than the value $0.90(2)$ from diagrammatic MC~\cite{prok08polaron} for a single impurity. Ansatz Eq.(\ref{e:wavefunction}) does not predict a transition, as it does not test for the formation of molecules.
In Fig.~\ref{f:residue}, the color coding reveals where molecular behavior is observed (yellow), and where the spectra show polaronic behavior (red to black). It can be seen that the critical interaction strength for the formation of molecules depends only weakly on the impurity concentration $x$.

In conclusion, we have observed Fermi polarons in a novel, attractive Fermi liquid of spin down impurity atoms immersed in a spin up Fermi sea. The energy and residue of this quasiparticle was determined and interactions between quasiparticles were found to be weak. Polarons thus emerge as the quasiparticles of a Landau Fermi liquid description of this strongly interacting Fermi mixture. To study first the impurity limit of $N+1$ interacting particles before dealing with the full $N+M$ many-body system will be a fruitful approach for other strongly correlated systems realized with cold atoms.
An intriguing question is how the limit of a weakly interacting polaron liquid containing few impurities connects to the physics of a hydrodynamic, balanced Fermi gas containing Cooper pair fluctuations above the critical temperature for superfluidity. In light of our findings, fermion pair condensation could be viewed as condensation of pairs of polarons with opposite spin. This is also suggested by the large normal state interaction measured by quasiparticle spectroscopy on the superfluid state~\cite{schi08gap}.

\begin{acknowledgments}
We would like to thank W. Ketterle, M. Randeria, S. Stringari, B. Svistunov, S. Todadri and W. Zwerger for helpful discussions, and Aviv Keshet for the computer control system. This work was supported by the NSF, an AFOSR-MURI, and the Alfred P. Sloan Foundation.
\end{acknowledgments}

\section{Supplemental Material: ``Observation of Fermi Polarons in a Tunable Fermi Liquid of Ultracold Atoms''}

In this supplemental material we state, starting with the variational Ansatz by Chevy, key properties of the polaron, such as the energy $E_\downarrow$ and the quasiparticle residue $Z$, and we calculate its RF spectrum using Fermi's Golden Rule. We connect this approach and its implication for finite impurity concentration to the results of Fermi liquid theory and the T-matrix formalism used in~\cite{punk07rf,veil08RFimbal,mass08polaron,schn09RF}. Furthermore, details are provided about the extraction of the quasiparticle residue $Z$.

\subsection{Polaron wavefunction, energy and quasiparticle residue}
We start with the hamiltonian for a dilute two component mixture of fermionic atoms interacting via the van-der-Waals potential $V(\vect{r})$~\cite{kett08varenna}. Thanks to the diluteness of the system, the potential is of short range $R$ compared to the interparticle distance $~1/k_F$, so $k_F R \ll 1$. Its Fourier transform $V(\vect{k})$ is thus essentially constant, $g_0$, below $k_F$ and rolls off to zero at a momentum on the order of $1/R \gg k_F$.
The many-body Hamiltonian for the system is then
\begin{equation}
    \hat{H} = \sum_{\vect{k},\sigma} \epsilon_\vect{k} c_{\vect{k}\sigma}^\dagger c_{\vect{k}\sigma} + \frac{g_0}{\cal V}\sum_{\vect{k},\vect{k}',\vect{q}} c_{\vect{k}+\frac{\vect{q}}{2} \uparrow}^\dagger c_{-\vect{k}+\frac{\vect{q}}{2}\downarrow}^\dagger c_{\vect{k}'+\frac{\vect{q}}{2}\downarrow} c_{-\vect{k}'+\frac{\vect{q}}{2}\uparrow}
\label{e:Hamiltonian}
\end{equation}
Here, the label $\sigma$ denotes the spin state $\uparrow$,$\downarrow$, $\epsilon_\vect{k} = \hbar^2 k^2 / 2m$, $\cal V$ is the volume of the system and the $c_{k\sigma}^\dagger$,$c_{k\sigma}$ are the usual creation and annihilation operators for fermions with momentum $\vect{k}$ and spin $\sigma$.
The trial wavefunction for the Fermi polaron with zero momentum proposed by F. Chevy in~\cite{chev06polaron} is
\begin{equation}
\left|\Psi\right> = \varphi_0 \left|\mathbf{0}\right>_\downarrow \left|FS\right>_\uparrow + \sum_{q<k_F}^{k>k_F} \varphi_{\mathbf{k}\mathbf{q}} c^\dagger_{\mathbf{k}\uparrow} c_{\mathbf{q}\uparrow}  \left|\mathbf{q}-\mathbf{k}\right>_\downarrow \left|FS\right>_\uparrow
\label{e:wavefunction}
\end{equation}
The energy is then minimized under variation of the parameters $\varphi_0$ and $\varphi_{\mathbf{k}\mathbf{q}}$, with the constraint of constant norm $\left<\Psi|\Psi\right> = \left|\varphi_0\right|^2 + \sum_{q<k_F}^{k>k_F} \left|\varphi_{\mathbf{k}\mathbf{q}}\right|^2 = 1$. That is, the quantity to minimize is
$\left<\Psi|\hat{H}|\Psi\right> - E_\downarrow \left<\Psi|\Psi\right>$.
The derivation can be found in~\cite{chev06polaron}, here we quote the result for the particle-hole excitation amplitudes $\varphi_{\vect{k}\vect{q}}$, the quasiparticle weight $\left|\varphi_0\right|^2 = Z$, and the energy $E_\downarrow$ due to addition of the down spin impurity:
\begin{eqnarray}
\varphi_{\vect{k}\vect{q}} &=& \varphi_0 \frac{1}{\cal V}\frac{f(E_\downarrow,\vect{q}) }{E_\downarrow - \epsilon_\vect{k} + \epsilon_\vect{q} - \epsilon_{\vect{q}-\vect{k}}} \label{e:phikq}\\
\frac{1}{\left|\varphi_0\right|^2} \equiv \frac{1}{Z} &=& \left(1 - \frac{\partial}{\partial E}\frac{1}{{\cal V}} \sum_{q<k_F} f(E,\vect{q})\right)_{E=E_\downarrow} \label{e:Z}
\end{eqnarray}
\begin{eqnarray}
E_\downarrow &=& \frac{1}{{\cal V}}\sum_{q<k_F} f(E_\downarrow,\vect{q})
\end{eqnarray}
These all depend on the function $f(E,\vect{q})$ with
\begin{eqnarray}
f^{-1}(E,\vect{q}) &=& \frac{1}{g_0} + \frac{1}{{\cal V}}\sum_{k>k_F} \frac{1}{\epsilon_\vect{k} - \epsilon_\vect{q}+\epsilon_{\vect{q}-\vect{k}} - E}
\end{eqnarray}
It is a measure of the interaction strength between spin up and spin down, modified by the presence of the spin up Fermi sea. As usual, $g_0$ can be replaced by the physically observable scattering length $a$ for collisions between spin up and down via~\cite{kett08varenna} $\frac{1}{g_0} = \frac{m}{4\pi\hbar^2 a} - \frac{1}{{\cal V}}\sum_\vect{k} \frac{1}{2\epsilon_\vect{k}}$.
\begin{eqnarray}
f^{-1}(E,\vect{q}) &=& \frac{m k_F}{2\pi^2\hbar^2}\left(\frac{\pi}{2 k_F a} - 1\right) + \\
 &&\frac{1}{{\cal V}} \sum_{k>k_F} \left(\frac{1}{\epsilon_\vect{k}-\epsilon_\vect{q} +\epsilon_{\vect{q}-\vect{k}} - E} - \frac{1}{2 \epsilon_k}\right) \nonumber
\end{eqnarray}
The integral in above expression is convergent and gives
\begin{eqnarray}
&&f^{-1}(E,\vect{q}) = \frac{m k_F}{2\pi^2\hbar^2}\left\{\frac{\pi}{2 k_F a} - 1 + I\left(\frac{E}{E_F},\frac{q}{k_F}\right)\right\} \\
&&I(\epsilon,y) =\int_1^\infty {\rm d}x \left(\frac{x}{2 y}\ln\left(\frac{2 x^2 + 2 x y - \epsilon}{2 x^2 - 2 x y - \epsilon}\right) -1\right) \nonumber
\end{eqnarray}
An analytic expression for the integral exists but does not provide additional insight.
The equation for $E_\downarrow$ becomes
\begin{eqnarray}
\frac{E_\downarrow}{E_F} =-2 \int_0^1 {\rm d}y \frac{y^2}{1 - \frac{\pi}{2 k_F a} -
 I\left(\frac{E_\downarrow}{E_F},y\right)}
\end{eqnarray}
This implicit equation for $E_\downarrow$ can be easily solved numerically. The result is shown as the dashed line in Fig. 3 of the main paper. Clearly, $E_\downarrow$ is negative due to the attractive interactions with the medium.
In the weakly interacting limit $1/k_F a \rightarrow -\infty$, we can neglect the integral in the denominator and immediately obtain
$E_\downarrow = \frac{2}{3}\frac{2 k_F a}{\pi} E_F = 4\pi\hbar^2 a n_\uparrow / m$, which is the mean field result~\cite{comb07polaron}.

The approach turns out to be equivalent to a T-Matrix description, as shown in~\cite{comb07polaron}. In that language, $f(E,\vect{q})$ is (up to a constant) the scattering amplitude in the medium (i.e. the vertex) for the scattering process with total energy and momentum $E$ and $q$ of the colliding particles.
$\Sigma(\vect{0},E) \equiv \frac{1}{{\cal V}}\sum_{q<k_F} f(E_\downarrow + E,\vect{q})$ is the self-energy at zero momentum and frequency $E/\hbar$. It is real in this approximation. The expression for the quasiparticle residue $Z$ of a {\it single} spin down impurity in Eq.~\ref{e:Z} is immediately seen to be equivalent to the well-known relation~\cite{veil08RFimbal}
\begin{equation}
    Z_{\downarrow}^{-1} = \left(1 -\frac{\partial}{\partial E} {\rm Re} \Sigma(k_{F\downarrow},E)\right)_{E = 0}
\end{equation}
for a spin down quasiparticle on top of a spin down Fermi sea, in the limit of vanishing Fermi momentum $k_{F\downarrow}$.

\subsection{RF spectrum from the variational Ansatz}

Fermi's Golden Rule allows us to directly predict the shape of the impurity RF spectrum. This topic has been studied in detail since the early days of RF spectrosocpy, and for the problem of highly imbalanced Fermi gases in~\cite{punk07rf,veil08RFimbal,mass08polaron,schn09RF}, among others. Chevy's wavefunction offers a simple way of calculating the RF spectrum of a single impurity.

The RF operator $\hat{V} = \hbar \Omega_R \sum_k c_{k,f}^\dagger c_{k,\downarrow} + h.c.$ promotes the impurity into the free final state $\left|f\right>$ (energy $E_f$) without momentum transfer~\cite{kett08varenna}. In the experiment, the final internal state is the second lowest hyperfine state of $^6$Li.
Fermi's Golden Rule for the impurity starting in state $\left|\Psi\right>$ is
\begin{equation}
    \Gamma(\omega) = \frac{2\pi}{\hbar} \sum_f \left|\left<f|\hat{V}|\Psi\right>\right|^2 \delta(\hbar\omega - (E_f - E_\downarrow)) \label{e:GoldenRule}
\end{equation}
Where $\omega$ is the RF offset from the bare atomic transition frequency between the internal states labeled by $\downarrow$ and $f$.
One possible final state is
$\left|0\right> \equiv \left|\vect{0}\right>_f \left|FS\right>_\uparrow$, i.e. a zero momentum particle in the final state plus a perfect Fermi sea of up spins, with energy $E_{\left|0\right>} = 0$ relative to the Fermi energy $E_F$ of the environment. Other possible final states are $\left|\vect{q},\vect{k}\right> \equiv \left|\vect{q}-\vect{k}\right>_f c^\dagger_{\vect{k}\uparrow} c_{\vect{q}\uparrow} \left|FS\right>_\uparrow$ with $q<k_F$ and $k>k_F$, i.e. a particle with momentum $\vect{q}-\vect{k}$ in the final state and a Fermi sea with a hole at $\vect{q}$ and an excited environment particle above the Fermi sea at $\vect{k}$. The energy of these states is $E_{\left|\vect{q},\vect{k}\right>} = \epsilon_\vect{k} - \epsilon_\vect{q} + \epsilon_{\vect{q}-\vect{k}}$ relative to the environment Fermi energy $E_F$.
The matrix elements are
\begin{eqnarray}
    \left<0\left|\hat{V}\right|\Psi\right> = \hbar \Omega_R\, \varphi_0 \nonumber \\
    \left<\vect{q},\vect{k}\left|\hat{V}\right|\Psi\right> = \hbar \Omega_R\, \varphi_{\vect{k}\vect{q}} \nonumber
\end{eqnarray}
This leaves us with two components in the RF spectrum:
\begin{eqnarray}
    \Gamma(\omega) &=& 2\pi\hbar\Omega_R^2\, \bigg(Z \delta(\hbar\omega+E_\downarrow) + \label{e:RFspectrum}\\
&&    \sum_{q<k_F}^{k>k_F} \left|\varphi_{\vect{k}\vect{q}}\right|^2 \delta(\hbar\omega +
    E_\downarrow - \epsilon_\vect{k} + \epsilon_\vect{q} - \epsilon_{\vect{q}-\vect{k}}) \bigg) \nonumber
\end{eqnarray}
The first part is a delta-peak shifted by the quasiparticle energy. As $E_\downarrow <0$, it is shifted to higher frequencies: The RF photon has to supply additional energy to transfer the impurity out of its attractive environment.
The weight of this peak is $Z$, the quasiparticle residue, allowing the experimental determination of $Z$ by simply integrating the area under the prominent peak.
Such a delta-peak is typically called "coherent", as a broadband excitation around this energy would not dephase over time. The second part of the spectrum is incoherent, it consists of a broad continuum of frequencies. Broadband excitations of this continuum would rapidly dephase, over a timescale given by the inverse width of the continuum.

\subsection{RF spectrum in Fermi Liquid theory}

This structure of the RF spectrum is a generic feature of quasiparticle spectra. In Fermi liquid theory, the propagator of a quasiparticle is approximated as a pole at energy $E(k)>0$ (relative to the ground state energy), lifetime $1/\gamma(k)$ and residue $Z$ plus an incoherent spectrum~\cite{veil08RFimbal,nozi97fermisystems}
\begin{equation}
    G_-^R(\vect{k},\omega) = \frac{Z}{\hbar\omega + E(k) + i \hbar\gamma(k)} + G_{-}^{R,\rm inc}(\vect{k},\omega)
    \label{e:Green}
\end{equation}
The spectral function is given by $A_-(\vect{k},\omega) = -\frac{1}{\pi} {\rm Im} G_-^R(\vect{k},\omega) = Z \frac{1}{\pi} \frac{\hbar\gamma(k)}{(\hbar\omega+E(k))^2 + \hbar^2\gamma(k)^2} + A_{-}^{\rm inc}(\vect{k},\omega) $ which tends to
\begin{equation}
A_-(\vect{k},\omega) = Z \delta(\hbar\omega + E(k)) + A_-^{\rm inc}(\vect{k},\omega)
\end{equation}
in the limit of small damping of the quasiparticle. $A_-(\vect{k},\omega)$ measures the probability that removing a particle with momentum $\vect{k}$ will cost an energy $-\hbar \omega$.
The RF spectrum in linear response is given by~\cite{mass08polaron}
\begin{equation}
\Gamma(\omega) = 2\pi\hbar\Omega_R^2 \sum_k A_-(\vect{k},\epsilon_\vect{k} - \mu - \hbar\omega) n_F(\epsilon_\vect{k}-\mu-\hbar\omega) \label{e:gammafermiliquid}
\end{equation}
where $\mu$ is the chemical potential of the quasiparticle and $n_F(x) = 1/(e^{\beta x} + 1)$ is the Fermi function that tends to $\theta(-x)$ at zero temperature.
This is intuitively understood: For a given momentum $\vect{k}$, the RF photon with energy $\hbar\omega$ has to provide the energy $\epsilon_{\vect{k}} - \mu$ (relative to the initial chemical potential $\mu$) to create a free particle in the final state. The rest, $\hbar \omega -\epsilon_{\vect{k}} + \mu$, is used to remove a particle from the initial state (probability $A_-(\vect{k},\epsilon_\vect{k} - \mu - \hbar\omega)$) if there exists such a particle (factor $n_F(\epsilon_\vect{k}-\mu-\hbar\omega)$). Eq.~\ref{e:gammafermiliquid} is equivalent to Fermi's Golden Rule Eq.~\ref{e:GoldenRule}~\cite{fett71}.
In the case where the spectral function is dominated by a quasiparticle peak, the spectrum becomes
\begin{eqnarray}
\Gamma(\omega) &=& 2\pi\hbar\Omega_R^2 Z \sum_k \delta(\epsilon_\vect{k} + E(k) - \mu - \hbar\omega)\nonumber \\ &&\times \;n_F(\epsilon_\vect{k}- \mu-\hbar\omega) + \Gamma^{\rm inc}(\omega)
\end{eqnarray}
Connecting to our case of a single quasiparticle with $\mu = E_\downarrow$ and $k = 0$, this directly gives
\begin{equation}
\Gamma(\omega) = 2\pi\hbar\Omega_R^2 Z \delta(\hbar\omega+E_\downarrow) + \Gamma^{\rm inc}(\omega)
\end{equation}
identical to the prediction via the trial wavefunction.

\subsection{Polaron spectral function}

To connect the single particle and the Fermi liquid description, we calculate the propagator $G^R_-(\vect{k},\omega)$ for the removal of a single spin down impurity from the wavefunction $\left|\Psi\right>$.
By definition,
\begin{eqnarray}
i G_-^R(\vect{k},t) &=& \left<\Psi\right| c^\dagger_{\vect{k}\downarrow}e^{i \hat{H} t/\hbar} c_{\vect{k}\downarrow} \left|\Psi\right>\theta(t)
\end{eqnarray}
Inserting a complete set of eigenstates, this gives
\begin{eqnarray}
i G_-^R(\vect{k},t) &=& \sum_f \left|\left<f\right|c_{\vect{k}\downarrow} \left|\Psi\right>\right|^2 e^{i E_f t/\hbar} \theta(t)
\end{eqnarray}
The state $c_{\vect{k}\downarrow} \left|\Psi\right>$ is void of any spin down impurity and has non-vanishing matrix elements only with either the unperturbed spin up Fermi sea, $\left|FS\right>_\uparrow$ (if $\vect{k} = \vect{0}$), or with particle-hole excitations $\left|\vect{q},\vect{k}'\right> = c_{\vect{k}'\uparrow}^\dagger c_{\vect{q} \uparrow} \left|FS\right>_\uparrow$ (in the case $\vect{k} = \vect{q}-\vect{k}'$). These matrix elements are $\varphi_0 = \sqrt{Z}$ and $\varphi_{\mathbf{k}'\mathbf{q}}$ resp., the corresponding energies $E_f = 0$ and $E_f = \epsilon_{\vect{k}'}-\epsilon_{\vect{q}}$ relative to $E_F$. So one has:
\begin{equation}
i G_-^R(\vect{k},t) =
\large(Z\delta_{\vect{k},\vect{0}} +\sum_{q<k_F}^{k'>k_F} \delta_{\vect{k},\vect{q}-\vect{k}'} \left|\varphi_{\vect{k}'\vect{q}}\right|^2 e^{i (\epsilon_{\vect{k}'}-\epsilon_{\vect{q}}) t/\hbar}\large) \theta(t)  \nonumber
\end{equation}
Finally, $G_-^R(\vect{k},\omega) = \frac{Z}{\hbar\omega +i\eta}\delta_{\vect{k},\vect{0}} + G_-^{R,\rm inc}(\vect{k},\omega)$ with infinitesimal $\eta>0$. This is just the Fermi liquid form of $G_-^R$ but for a single quasiparticle with zero momentum ($E(0) = 0$ in (\ref{e:Green})), as described by $\left|\Psi\right>$.
The spectral function is
\begin{eqnarray}
A_-(\vect{k},\omega) &=& Z \delta(\hbar\omega)\delta_{\vect{k},\vect{0}} + \nonumber \\
&&\sum_{q<k_F}^{k'>k_F} \delta_{\vect{k},\vect{q}-\vect{k}'} \left|\varphi_{\vect{k}'\vect{q}}\right|^2 \delta(\hbar\omega +\epsilon_{\vect{k}'}-\epsilon_{\vect{q}}) \nonumber
\end{eqnarray}
With Eq.~\ref{e:gammafermiliquid} this exactly gives the RF spectrum of Eq.~\ref{e:RFspectrum}.

\subsection{Calculation of the incoherent background}
Using Eq.~\ref{e:phikq}, we can write the incoherent part of the spectrum as:
\begin{eqnarray}
   \Gamma^{\rm inc}(\omega) \equiv 2\pi\hbar\Omega_R^2 \sum_{q<k_F}^{k>k_F} \left|\varphi_{\vect{k}\vect{q}}\right|^2 \delta(\hbar\omega +
    E_\downarrow - \epsilon_\vect{k} + \epsilon_\vect{q} - \epsilon_{\vect{q}-\vect{k}}) \nonumber \\
    = \frac{2\pi \Omega_R^2}{\hbar}  \frac{Z}{\omega^2} \frac{1}{{\cal V}^2} \sum_{q<k_F}^{k>k_F} f(E_\downarrow,\vect{q})^2 \delta(\hbar\omega +
    E_\downarrow - \epsilon_\vect{k} + \epsilon_\vect{q} - \epsilon_{\vect{q}-\vect{k}}) \nonumber
\end{eqnarray}
The integral over $\vect{k}$ exists in analytic form:
\begin{eqnarray}
    \frac{1}{\cal V}&\sum_{k>k_F} \delta(\hbar\omega +
    E_\downarrow - \epsilon_\vect{k} + \epsilon_\vect{q} - \epsilon_{\vect{q}-\vect{k}})
    = \nonumber \\
&\frac{m k_F}{8\pi^2\hbar^2} K(\frac{\hbar\omega + E_\downarrow}{2 E_F},\frac{q}{k_F})& \nonumber \\
\text{with }
&K(\epsilon,y) =\begin{cases}
     \frac{y_+^2 - y_-^2}{y} & \text{for } y_- > 1\,, \\
     \frac{y_+^2 - 1}{y} & \text{for } y_- < 1 < y_+\,, \\
     0 & \text{for } 1 > y_+\, .
\end{cases}& \nonumber \\
\text{and }
&y_\pm = \pm \frac{y}{2} + \sqrt{\frac{y^2}{4} + \epsilon} &
\end{eqnarray}
The incoherent spectrum is then
\begin{eqnarray}
\Gamma^{\rm inc}(\omega) = \pi \Omega_R^2 \frac{Z E_F}{\hbar\omega^2}
\int_0^1 {\rm d}y \frac{y^2 K\left(\frac{\hbar\omega + E_\downarrow}{2 E_F},y\right)}{\left(1 - \frac{\pi}{2 k_F a} -
 I\left(\frac{E_\downarrow}{E_F},y\right)\right)^2} \label{e:Gammainc}
\end{eqnarray}
One can check that the total spectrum obeys the sum rule
\begin{equation}
    \int_{-\infty}^{\infty} {\rm d}\omega \Gamma(\omega) = 2\pi\hbar\Omega_R^2
\end{equation}
and in particular that the total weight of the incoherent background is proportional to $1 - Z$, which is not obvious from the form in Eq.~\ref{e:Gammainc}.
For RF frequencies close to threshold $\hbar\omega + E_\downarrow \ll 2E_F$, the hole momentum $\vect{q}$ and the particle momentum $\vect{k}$ must be close to each other to fulfill energy conservation, i.e. they have to be close to the Fermi momentum. The double sum over $\vect{q}$ and $\vect{k}$ thus gives a phase space suppression on the order of $(\hbar \omega+E_\downarrow)^2$, i.e. the spectrum starts like $(\hbar \omega+E_\downarrow)^2/\omega^2$. This is in contrast to the dissociation spectrum of a molecule of binding energy $E_B$, where the density of states above threshold gives a spectrum proportional to $\sqrt{\hbar \omega+E_B}/\omega^2$.
For large RF energies, large particle momenta $\vect{k}$ are involved, the suppression due to the Fermi sea becomes negligible and the spectrum behaves like $\sqrt{\hbar\omega+E_\downarrow}/\omega^2$, as for a molecule of binding energy $E_\downarrow$. This is natural as for large momenta, we are probing short-range physics which involves at most two particles, a spin up environment atom and the impurity. In particular, at RF energies $\hbar\omega \gg E_\downarrow$, we recover the $\omega^{-3/2}$ behavior of the RF spectrum that is universal for short-range interactions.

\subsection{RF Spectrum of a finite concentration of impurities}

Since polarons are found to be weakly interacting, they will form a Fermi sea filled up to the impurity Fermi momentum $k_{F\downarrow}$. The fact that the dispersion of polarons $E_(k) = \frac{m}{m^*}\epsilon_\vect{k}$ differs from that of a free particle due to the effective mass $m^*\ne m$ leads to broadening of the RF spectra. The RF photon has to supply the difference in kinetic energies $(1-\frac{m}{m^*})\epsilon_\vect{k}$ between the initial and the final state, with a maximal shift $(1-\frac{m}{m^*})\hbar^2 k_{F\downarrow}^2/2m$.
The spectral shape is easily obtained: The spectral function at momentum $\vect{k}$ will be dominated by polarons that occupy that momentum state. The coherent part of the spectral function is thus $A^{\rm coh}_-(\vect{k},\omega) = Z\delta(\hbar\omega+E(k))$ with $E(k) = -\hbar^2 k^2 /2m^* = -\frac{m}{m^*}\epsilon_\vect{k}$ relative to the impurity Fermi energy.
The coherent part of the spectrum then becomes
\begin{eqnarray}
\Gamma^{\rm coh}(\omega) = 2\pi\hbar\Omega_R^2 \sum_k A^{\rm coh}_-(\vect{k},\epsilon_k - E_\downarrow-\hbar\omega)
\nonumber
\end{eqnarray}
where the sum extends up to the impurity Fermi momentum $k_{F\downarrow}$. With the free, 3D density of states $\rho(\epsilon)$, this is
\begin{eqnarray}
\Gamma^{\rm coh}(\omega) &=& 2\pi\hbar\Omega_R^2 \int_0^{E_{F\downarrow}} {\rm d}\epsilon \, \rho(\epsilon) Z\delta(\epsilon - E_\downarrow-\hbar\omega-\frac{m}{m^*}\epsilon) \nonumber \\
&=& 2\pi\hbar\Omega_R^2 \frac{Z}{1-\frac{m}{m^*}}\rho\left(\frac{\hbar\omega+E_\downarrow}{1-\frac{m}{m^*}}\right) \times\nonumber \\
&&\theta\left((1-\frac{m}{m^*})E_{F\downarrow}-\hbar\omega-E_\downarrow\right)
\end{eqnarray}
This coherent part of the spectrum starts at the polaron ground state energy $\hbar\omega = |E_\downarrow|$, then grows like a square root and jumps to zero when $\hbar\omega-|E_\downarrow| = (1-\frac{m}{m^*})E_{F\downarrow}$. On resonance, where $m^* \approx 1.2$, this occurs at $\hbar\omega-|E_\downarrow| = 0.2 x^{2/3}E_{F\uparrow} \approx 0.04 E_{F\uparrow}$ for $x = 0.1$. This is still smaller than the Fourier width of the RF pulse used in the experiment of about $0.1 E_F$. The size of the jump is given by $2\pi\hbar\Omega_R^2 \frac{Z}{1-\frac{m}{m^*}} \rho(E_{F\downarrow})$ and reflects the impurity Fermi surface in the RF spectrum.
This behavior of the coherent part of the spectrum was found in~\cite{punk07rf} and was discussed recently in~\cite{schn09RF}. It is intriguing that the sharpness of the Fermi surface and its discontinuity should, at least in principle, be observable in the RF spectrum.

\subsection{Determination of $Z$ from experimental spectra}
\begin{figure}
\includegraphics[width=2.75in]{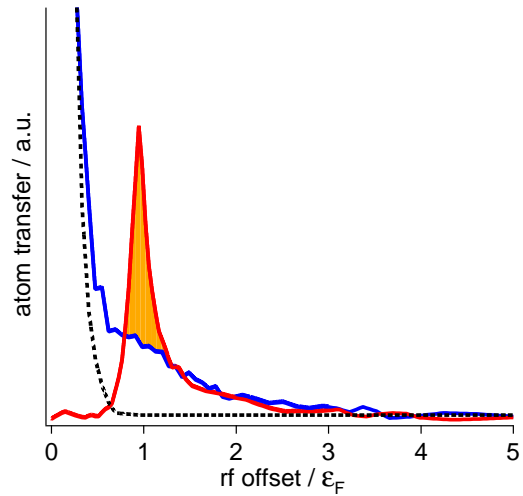}
    \caption{Determination of the quasiparticle residue $Z$. Impurity spectrum (red), environment spectrum (blue) and spectral response of non-interacting atoms (black dashed), folded over from negative RF offsets.}
    \label{f:Zextraction}
\end{figure}

In order to extract the quasiparticle residue $Z$, we determine the area under the peak that is not matched by the environment's response and divide by the total area under the impurity spectrum (see spectrum in the inset of Fig. 4 in the main body of the paper). Due to the Fourier width of the probe pulse, the strong response of the environment around zero RF offset (the resonance for non-interacting atoms) adds some weight to the environment background at the position of the polaron peak. To remove this effect in the determination of $Z$, the part of the environment's response at negative frequency offset is folded towards the positive side (dashed line in Fig.~\ref{f:Zextraction}) and subtracted from the environment spectrum. As it turns out, this procedure changes the value for Z by less than 5\% for all spectra in Fig. 2 of the main paper.

\begin{thebibliography}{24}
\bibitem{footpauli}
We do not observe Pauli blocking effects proposed in~\cite{mass08polaron}.
\bibitem{footfinalstate}
The next order correction due to incoherent scattering, where $a_{\rm fe}$ is replaced by the full real part of the scattering amplitude, is of order $(k_F a_{\rm fe})^2 \approx 5\%$ and thus negligible.
\end{thebibliography}
\end{document}